\documentclass[twocolumn,preprintnumbers,amsmath,amssymb]{revtex4}
\usepackage{graphicx}
\usepackage{dcolumn}
\usepackage{bm}
\begin{document}
\preprint{NStaley}
\title{Electric-field driven insulating to conducting transition in a mesoscopic quantum dot lattice }
\author{Neal E. Staley}
\email{nstaley@mit.edu}
\author{Nirat Ray}
\author{Marc A. Kastner}

\affiliation{Department of Physics, Massachusetts Institute of Technology, Cambridge, Massachusetts 02139, U.S.A.}
\author{Micah P. Hanson}
\author {Arthur C. Gossard}
\affiliation{Materials Department, University of California, Santa Barbara, California 93106-5050, U.S.A}
\date{\today}
\begin{abstract}
We investigate electron transport through a finite two dimensional mesoscopic periodic potential, consisting of an array of lateral quantum dots with electron density controlled by a global top gate. We observe a transition from an insulating state at low bias voltages to a conducting state at high bias voltages. The insulating state shows simply activated temperature dependence, with strongly gate voltage dependent activation energy.  At low temperatures the transition between the insulating and conducting states becomes very abrupt and shows strong hysteresis.  The high-bias behavior suggests underdamped transport through a periodic washboard potential resulting from collective motion.  
\end{abstract}
\maketitle

There has been great interest in understanding the motion of charge carriers in artificial periodic potentials with mesoscopic periods\cite{esaki1970,thouless1982,peeters1992}.  In particular, one might better understand transport in general by controlling the energy scales of importance:  For quantum transport in such systems the important energies are the on-site excitation and Coulomb charging energies and the inter-site tunneling matrix element.  Control of these energies has been demonstrated in single lateral quantum dots connected by tunneling to leads, leading to insights into the Kondo effect\cite{goldhaber1998}, for example; similar insights into the Hubbard model might emerge from experiments on arrays of lateral quantum dots\cite{byrnes2008,nictua2013,singha2011}.  For classical transport modeled as charge diffusion through a tilted washboard potential, applicable to a wide variety of experimental systems\cite{Tinkham,gruner1988,costantini1999}, the energy scale is the height of the potential barrier between sites.  In addition to the general question of whether quantum or classical transport dominates, artificial periodic potentials may provide insights into the extensive work on self-assembled arrays of semiconductor nanocrystals useful for opto-electronic devices\cite{murray1995,leatherdale2000}.

Many years ago, Duru{\"o}z \textit{et al.} reported switching and hysteresis in an array of 200 by 200 lateral quantum dots in GaAs/AlGaAs heterostructures\cite{duruoz1995}.  Subsequent experiments have been unable to reproduce these effects\cite{dorn2004,goswami2012}, raising the possibility that the observed hysteresis results from the leakeage current between the gate and dots observed by  Duru{\"o}z \textit{et al.}  In this Letter we report detailed measurements of the current through a 10 by 10 array of lateral quantum dots with a period of 340 nm.  Like Duru{\"o}z \textit{et al.}, we find a hysteretic transition from a high resistance state at low bias to a low resistance state at high bias which is strongly tuned by magnetic field.  Unlike previous measurements our devices have immeasurably small leakage between the gate and the dot array. We have studied the temperature, magnetic field, and gate-voltage (V$_g$) dependence of the transition in great detail.  We find evidence that the important energy scale for transport within the high-resistance state is the barrier height between quantum dots. However, the low-resistance state is quite unusual.  While some of the features we observe are predicted by the simulations of the motion of a classical charged particle in a washboard potential with finite pinning, the observation of dramatic hysteresis in such a small array suggests that collective motion of the charge carriers may be important.

\begin{figure}
\includegraphics[ scale =1]{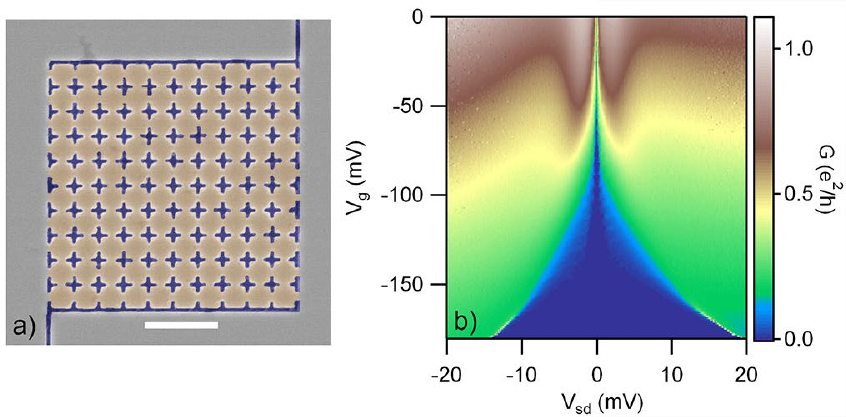}
\caption{(Color online) a) False color scanning electron microscope (SEM) image of a simlar device to the one reported in this paper, the scale bar is 1 $\mu$m. The etched lines are false colored blue, while the square quantum dot array is false colored orange.  b) Two dimensional color contour plot showing the differential conductance as a function of source drain bias and gate voltage at 50 mK.}
\end{figure}

We have fabricated a square array of dots, each with a lithographic area of 0.09 $\mu$m$^2$, and narrow barriers between them. We have utilized reactive ion etching to define the dots in AlGaAs/GaAs heterostructures, with mobility of 6.4x10$^5$cm$^2$/Vs and carrier density of 2.2x10$^{11}$cm$^{-2}$.  We stop the etching after penetrating through the dopant layer but prior to reaching the GaAs active region. A scanning electron microscope image of a similar device is shown in Fig. 1a. To tune the total charge-carrier density we utilize a global top gate, which is isolated from the quantum dots by 100 nm of evaporated SiO$_2$. The data presented here is measured using a two terminal geometry, with contact resistances into the array of $<$5k$\Omega$, using both AC and DC techniques in a dilution refrigerator with room temperature RF filtering and low temperature microwave filtering.

\begin{figure}
\includegraphics[ scale =1]{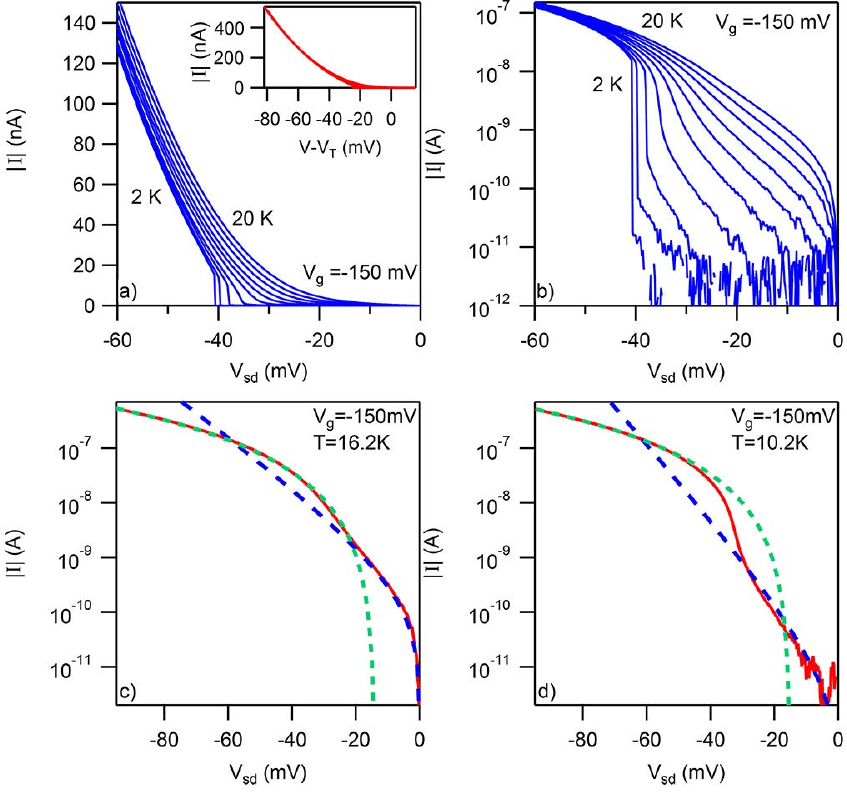}
\caption{(Color Online) Current vs voltage at temperatures between 2 and 20K at V$_g$=-150 mV on a linear plot a) and semi-log plot b).  For each successive trace the temperature is increased by 2K.  The inset in a) shows the same traces plotted vs (V-V$_T$) c) and d ) are semi-log plots of the magnitude of the current vs. source drain bias  (solid red curve), with fits to the high bias power law (dashed green) and the low exponential (dashed blue) at two specific temperatures.}
\end{figure}

\begin{figure}
\includegraphics[ scale =1]{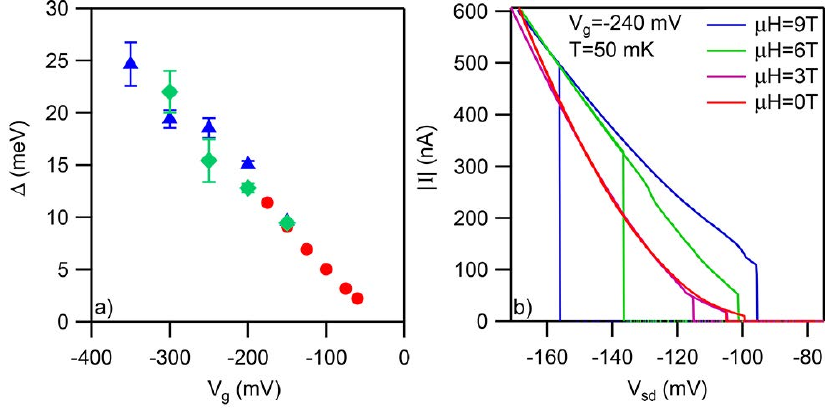}
\caption{(Color Online) a) The gate voltage dependence of the single particle gap.  The single particle gap in zero magnetic field was extracted from the temperature dependence of the zero bias conductance (red circles) and from the fits to the I-V curves (blue triangles).  The single particle gap at 9T extracted from I-V curves is shown as green diamonds.  b) Plot showing current voltage (I-V) hysteresis loops taken atV$_g$=-240 mV for applied magnetic fields from 0T to 9T.  }
\end{figure}

Measurements of our device at low temperatures reveal a non-conducting to conducting transition driven by source drain bias (V) that has a strong dependence on the gate voltage (V$_g$). In Fig. 1b we show the differential conductance as a function of V$_g$ and V at 50 mK. At this temperature, for positive gate voltages, which correspond to adding electrons into the array, we see a component with finite zero-bias conductance, but with a zero-bias anomaly --a rapid increase of conductance with small bias.  As we decrease the gate voltage, which begins to deplete the array, the zero bias anomaly broadens and evolves, near V$_g$= -75 mV, into a non-conducting state at V=0. At more negative V$_g$ the current-voltage (I-V) characteristic has a distinct threshold for conduction and a power-law dependence above threshold. As V$_g$ is made more negative the threshold becomes abrupt.  Very similar behavior is seen for negative and positive V. Similarly abrupt transitions from an insulating to conducting state in quantum dot arrays have been previously observed and attributed to leakage current between the top gate and the array\cite{duruoz1995}. In our experiment, we observe no measurable gate leakage with applied gate voltages up to -1.1V. 

To better characterize the insulating-to-conducting transition we have measured its temperature dependence. Shown in Fig. 2a are a series of I-V traces taken at temperatures from 2 to 20 K at V$_g$= -150 mV. In the log plot of the same data (Fig. 2b), one can clearly see  both the low bias non-conducting state and the high bias conducting state, as well as the transition between the two, which becomes abrupt below 10 K. At temperatures below 2K for this V$_g$ the current in the low bias state becomes immeasurable.  We see that at sufficiently high temperatures there is finite conductivity at zero bias, which freezes out as we decrease the temperature, suggesting that the non-conducting state is a gapped insulator. In contrast, the high bias behavior shows only a weak temperature dependence of the threshold voltage up to 20K.  Interestingly, the high bias curves collapse into a single trace by shifting the voltage axis by the threshold voltage as shown in the inset of Fig. 2a.  The slight deviations from a single collapsed trace seen in at very high bias might be due to electron self heating.  

\begin{figure*}
\includegraphics[ scale =1]{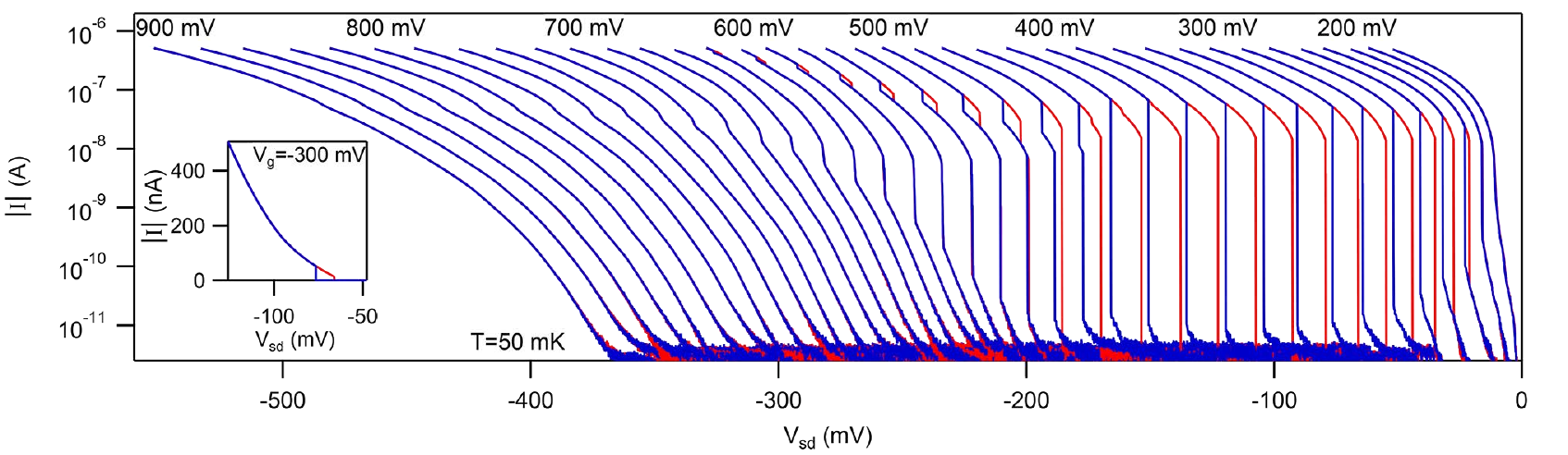}
\caption{(Color Online) Semi-log plot of the magnitude current vs source drain bias at 50 mK for hysteresis loops taken every 20 mV for gate voltages ranging from -160mV to -900mV. Note the strong hysteresis between gate voltages of -220 mV to -580 mV, also note the splitting of the switching behavior which could indicate the formation of two distinct domains. Inset shows a single hysteresis loop taken at V$_g$=-300 mV plotted in a linear scale.}
\end{figure*}

We find that a simple model describes the I-V characteristic of the array:  At low V, the current is consistent with that of thermally activated single particles overcoming a barrier,
\begin{equation}
I=VG_0e^{\frac{-(\Delta-eVx/L)}{k_BT}}
\end{equation}
with an activation energy $\Delta$ that is reduced by the field  across the barrier Vx/L, where L is the length of the array and x is width of the barrier.  Well above the threshold voltage, V$_T$, the current follows a power law, 
\begin{equation}
I=C(V-V_T)^\alpha
\end{equation}
Fits to both components are shown in Fig. 2c and d for temperatures of 16.2 K and 10.2K, respectively, using the same values of G$_0$, x, $\Delta$, C and $\alpha$.  From the low bias behavior we can extract the characteristic length, x/L. For this array and gate voltage x/L=0.12$\pm$0.01, corresponding to x = 410$\pm$50 nm.  As can be seen from Fig.2c,d, the exponentially voltage-dependent process should dominate, were it present at high bias voltages, but this is not observed. Similarly, the power-law fit to the high-bias current, overestimates the current at low biases.  Thus the two transport processes do not add, suggesting that they compete with each other. 

We measure $\Delta$ from the temperature dependence of the zero-bias conductance and fits to the I-V characteristics. For low gate voltages and low magnetic fields where the device is weakly insulating, the zero-bias conductance measurements are used, while for high gate voltages the I-V characteristics are used.  $\Delta$ is strongly V$_g$ dependent, increasing from 2.2 meV at a V$_g$= -60 mV, to 24.6 meV at a V$_g$= -350 mV as shown in Fig 3a.  The activation energy shows negligible dependence on an externally applied magnetic field up to 9T as seen in Fig. 3a.  

Perhaps the most striking feature of our data is the abruptness and the hysteresis associated with the transition between the insulating and conducting states observed at temperatures below 10 K. At base temperature, this abrupt transition is extremely sharp, with current jumps of more than three orders of magnitude observed at zero magnetic field.  In a magnetic field the insulating state is strongly stabilized, making the hysteresis much larger, as shown in Fig. 3b, showing a 5 orders of magnitude change between consecutive voltage points.  What is remarkable is that the magnetic field both stabilises the insulating state and increases the conductance of the conducting state.  

The abrupt transition between the insulating and conducting states exists over a large range of gate voltages at base temperature as shown in Fig. 4.  This transition becomes hysteretic near a gate voltage of -220 mV at zero magnetic field.  As we further deplete the array by making V$_g$ more negative, the abrupt current step rapidly saturates in amplitude until high gate voltages where we observe additional transitions and a gradual decrease in hysteresis amplitude. For V$_g$ more negative than $\sim$-700 mV, no transitions are observed, and the power law exponent becomes constant at $\sim$5.6. We expect that at such large negative V$_g$ the array is largly depleted of electrons.

To interpret our results we first note that the activation energies we observe, especially at V$_g$ more negative than  $\sim$-200 mV are much too large to be associated with the charging energy of a single dot, which we estimate, from the lithographic dimensions, to be $\sim$3.5 meV.  The tunneling matrix element and on-site excitation energies are expected to be even smaller.  This and the dramatic increase in the activation energy with more negative V$_g$ strongly suggest that the high-resistance state is limited by thermal activation over the barrier between sites in the array.  Thus, models like that of Middleton and Wingreen\cite{middleton93}, in which the transport is limited by the Coulomb charging energy would seem to be excluded.  

We next note that the multiple transitions seen, for example, in Fig. 4, suggest that the sample contains a number of domains, likely nucleated at disorder by disorder within the array, that undergo the transition independently, once the barrier height is large.  We have also made true 4 terminal measurements on a much larger 19x39 array, and find qualitatively similar results.  We therefore infer that in our geometry the high bias field tilts the washboard potential, reducing the barrier height for carrier motion down-field but leaving the barrier height in the perpendicular direction unchanged, leaving effectively ten rows of dots.  It is therefore reasonable to assume that the transition occurs when the washboard potential is tilted by a critical field.  We find that the threshold is roughly linear in gate voltage and the activation energy (see Fig. 3) is linear in gate voltage, so the transition appears to occur when the activation energy is reduced to a critical value near 10 meV.  Disorder will then cause different rows to undergo transitions at somewhat different fields.  We note, however, that there are far fewer transitions than rows of dots, indicating that the transitions in adjacent rows are somewhat correlated.  

Abrupt conductance changes have been previously observed in a wide variety of experimental systems resulting from correlated\cite{hall1988,lindelof1981}, or non-equilibrium\cite{ovadia2009} physics. In particular, for strongly temperature dependent systems, thermal runaway can cause dramatic increases in the conductivity as the source drain bias is increased\cite{ovadia2009}. In our quantum dot arrays the measured activation energies are so large that a temperature rise of over 15 K would be required to explain the switching at low temperature. One usually observes a considerable conductance increase prior to run-away, and we see no such precursor at low T and large negative V$_g$.  Furthermore, the large change in hysteresis observed at high magnetic fields, which does not influence the observed single particle gap within the insulating state but dramatically increases the power dissipated in the device in the conducting state strongly suggests this is not thermal runaway.  It is well-known that impact ionization or avalanche breakdown can give rise to abrupt increases in current, as well as hysteresis. While the electric fields where we observe the transitions are rather large, $>$150 V/cm, we would not expect the observed temperature or magnetic field dependence of the step transitions  and hysteresis in such a small system, were such mechanism responsible. 

There has been a good deal of theoretical work on the motion of single classical charge carriers in an  infinite washboard potentia\cite{RiskenBook,lindenberg2007,marchenko2014}.  Our system can be thought of as a small section of such a potential.  The simulations show that at low field the carriers are pinned in the potential wells in an insulating ``stationary'' state, with an activation energy determined by the barrier height between wells, while above a characteristic threshold field a fraction of carriers are  free to traverse the washboard in a conducting ``running'' state.  Reference\cite{marchenko2014} claims that there is hysteresis when the field is lowered.  However, it seems likely that the hysteresis in the simulations is the result of the infinite size of the system.  It is difficult to imagine that for a system as small as ours, the running state could be sustained for the many minutes necessary to yield the observed hysteresis.

Many of the features we see are reminiscent of the behavior of charge-density waves (CDWs), which are described by motion of a classical particle in a tilted washboard potential.  In that case hysteresis arises from the momentum of the collective state, once the threshold for its motion is overcome.  The power law I-V characteristic observed at high biases, with an exponent of 2.3 differs from the CDW exponent of  $\frac{3}{2}$\cite{gruner1988}, possibly due to the finite sample size or non-sinusoidal potential.  Due to the hysteresis and other features, we suggest that the high-field state results from collective motion of the carriers in the periodic potential.  Experiments are planned to search for the narrow-band noise anticipated for such motion.

\begin{acknowledgements}
We acknowledge useful discussions with Prof. Leonid S. Levitov, Prof. Ivan Marchenko and Dr. William D. Oliver. The work at MIT was supported by the NSF under Grant DMR-1104394, the work at UC Santa Barbara was supported by the NSEC Program of the NSF under Grant No. PHY-0117795.  NR acknowledges support from the Schlumberger Foundation through the Faculty for the Future Fellowship Program.
\end{acknowledgements}


\end{document}